\documentclass[aps,prstab,superscriptaddress,floatfix,twocolumn,amsfonts,amsmath,amssymb,amsart,longbibliography]{revtex4-2}

\usepackage{graphicx} % include figure files
\usepackage{dcolumn} % align table columns on decimal point
\usepackage{bm} % bold math
\usepackage{color}

\usepackage[colorlinks, breaklinks=true,linkcolor=red, citecolor=blue, linktocpage=true]{hyperref}

\begin{document}
	
	\title{Micromagnets dramatically  enhance effects of viscous hydrodynamic
		flow in two-dimensional electron fluid}

	\author{Jack N. Engdahl}
	\thanks{j.engdahl@student.unsw.edu.au}
	\affiliation{School of Physics, University of New South Wales, Sydney 2052, Australia}
	
	\author{Ayd\i n Cem Keser}
	\thanks{a.keser@unsw.edu.au}
	\affiliation{School of Physics, University of New South Wales, Sydney 2052, Australia}
	\author{Oleg P. Sushkov}
	\affiliation{School of Physics, University of New South Wales, Sydney 2052, Australia}
	
	\date{\today}
	\begin{abstract}
		{ The hydrodynamic behavior of electron fluids in a certain range of
			temperatures and densities is well established in graphene and in 2D
			semiconductor heterostructures. The hydrodynamic regime is intrinsically
			based on electron-electron interactions, and therefore it provides a
			unique opportunity to study electron correlations. Unfortunately, in simple longitudinal resistance measurements, the relative contribution of hydrodynamic effects to
			transport is rather small, especially at higher temperatures.
			Viscous hydrodynamic effects are  masked by impurities, interaction with
			phonons, uncontrolled boundaries and ballistic effects. This essentially
			limits the accuracy of measurements of electron viscosity. Fundamentally, what causes viscous friction in the electron fluid
			is the property of the flow called vorticity. In this paper, we propose to use
			micromagnets to increase the vorticity by orders of magnitude. Experimental
			realization of this proposal will bring electron hydrodynamics to a
			qualitatively new precision level, as well as opening a new way to characterize
			and externally control the electron fluid.
		}

	\end{abstract}

	\maketitle
	\section{Introduction}
	Viscous hydrodynamic flow of the electron fluid, first postulated by Gurzhi over half a century
	ago,\cite{gurzhi_hydrodynamic_1968}
	is now well established experimentally.~\cite{polini_viscous_2020}
	In the hydrodynamic regime, the momentum conserving electron-electron scattering
	timescale is much shorter than the momentum relaxation timescale related to impurities
	and phonons, allowing neighboring electrons to establish local thermal equilibrium such that the flow behaves as a fluid.  
	This condition requires samples of very high purity, therefore 
	the theoretical predictions of the hydrodynamic regime in the 1960s~\cite{gurzhi_hydrodynamic_1968,gurzhi_minimum_1963} could only be realized experimentally after the advent of high mobility semiconductors~\cite{de_jong_hydrodynamic_1995} and graphene.~\cite{lucas_hydrodynamics_2018,polini_viscous_2020}
	{
		The opposite side of the story is that the relative contribution of viscous hydrodynamic effects to
		transport is bounded even in modern high mobility samples, which drastically limit the precision of all the existing measurements of viscosity. In this work, we propose a way to overcome
		this limitation.
	}
	
	The hydrodynamic nature of the electron flow manifests itself in a number of phenomena that have been demonstrated in recent studies. While the hydrodynamic regime was first thought to be experimentally realized in 1995 in Ref.\cite{de_jong_hydrodynamic_1995}, this interpretation of the results was challenged in the same year in Ref.\cite{gurzhi_electron-electron_1995}, as the requirements for viscous hydrodynamics in a two-dimensional material differs from that in three-dimensions. Reliable evidence of the hydrodynamic regime was first presented two decades later in Refs.\cite{bandurin_negative_2016, moll_evidence_2016, alekseev_negative_2016}. Notable consequences of hydrodynamic behavior are: the decrease of resistance with increasing temperature known as the Gurzhi effect,\cite{de_jong_hydrodynamic_1995,krishna_kumar_superballistic_2017,bandurin_fluidity_2018,keser_geometric_2021, gusev_viscous_2018,ginzburg_superballistic_2021}
	Poiseuille flow profiles, \cite{sulpizio_visualizing_2019,ku_imaging_2020,ella_simultaneous_2019}
	negative non-local resistance and formation of whirlpools, \cite{govorov_hydrodynamic_2004,kaya_nonequilibrium_2013,bandurin_negative_2016,braem_scanning_2018} 
	Hall viscosity, \cite{gusev_viscous_2018,berdyugin_measuring_2019} negative magnetoresistance,~\cite{gusev_viscous_2018-1,shi_colossal_2014,keser_geometric_2021,mani_size-dependent_2013, alekseev_negative_2016,wang_hydrodynamic_2022} the violation of the Wiedemann-Franz
	law, \cite{crossno_observation_2016,gooth_thermal_2018,lucas_electronic_2018, jaoui_thermal_2021, ahn_hydrodynamics_2022}
	anomalous scaling of resistance with channel width, \cite{gooth_thermal_2018,moll_evidence_2016}, resonant photoresistance in magnetic field~\cite{dai_observation_2010,hatke_giant_2011,bialek_photoresponse_2015,alekseev_transverse_2019,wang_hydrodynamic_2022} and quantum-critical dynamic conductivity.~\cite{gallagher_quantum-critical_2019} Furthermore, many novel phenomena have been proposed, including the elimination of  Landauer-Sharvin resistance,~\cite{stern_spread_2021} anisotropic fluids,~\cite{varnavides_electron_2020,link_out--bounds_2018} dynamo effect in the electron-hole plasma~\cite{galitski_dynamo_2018} and most excitingly the prospect of hydrodynamic spin transport~\cite{tatara_hydrodynamic_2021,doornenbal_spinvorticity_2019,matsuo_spin_2020} inspired by experiments on liquid mercury.~\cite{takahashi_spin_2016,takahashi_giant_2020}
	{ So far almost all experiments in the field have been aimed at a qualitative
		demonstration of various hydrodynamic effects. Extraction of  quantitative
		information has been hindered by the weakness of the contribution of
		hydrodynamics effects to electron transport. The quantitative information is
		important for understanding correlation effects and it is  essential for
		future technological applications of the electron fluid.}

	%%%%%%%%%%%%%%%%%%%%%%%%%%%%%%%%%%%%%%%%%%%%%%%%%
	\begin{figure}[ht!]
		\includegraphics[width=0.5\textwidth]{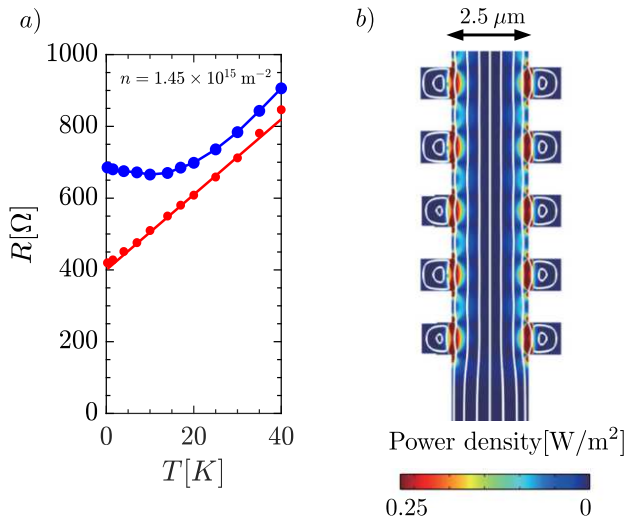}
		\caption{a) Resistivity of the GaAs device from Ref.~\cite{keser_geometric_2021}
			versus temperature. The red line represents the ohmic resistance
			and the  blue line represents the total (ohmic plus viscous) resistance.
			b) The map of the viscous dissipation
			(proportional to vorticity) in the device (drawn to scale). Vorticity is concentrated around sharp corners and practically absent in the bulk.
		}
		\label{AD}
	\end{figure}
	%%%%%%%%%%%%%%%%%
	
	%Recent studies reveal the hydrodynamic nature of electron flow, with the demonstration of
	%the Gurzhi effect\cite{de_jong_hydrodynamic_1995,krishna_kumar_superballistic_2017,gusev_visc%ous_2018-1},
	%Poiseuille flow profiles \cite{sulpizio_visualizing_2019,ku_imaging_2020},
	%negative local resistance \cite{govorov_hydrodynamic_2004,bandurin_negative_2016},
	%Hall viscosity \cite{gusev_viscous_2018,berdyugin_measuring_2019}, the violation of the %Wiedemann-Franz
	%law \cite{crossno_observation_2016,gooth_thermal_2018},
	%anomalous scaling of resistance with channel width %\cite{gooth_thermal_2018,moll_evidence_2016},
	%and experimental detection of the transition from the ballistic to the hydrodynamic
	%regime \cite{bandurin_fluidity_2018}.
	
	In the hydrodynamic regime, the flow of electrons is characterized by the viscosity of the electron fluid.
	Hydrodynamic flow of electrons may be analyzed using the methods of classical fluid
	mechanics, particularly solving the Navier-Stokes equations. \cite{landau_fluid_2013}
	While the flow is classical, the viscosity of the electron fluid is determined by
	electron-electron scattering that is purely quantum mechanical. Therefore, precise measurements of
	viscosity provides a new way to study electron-electron correlations.~\cite{alekseev_viscosity_2020}
	The most precise  measurements of the viscosity of electron fluid in 2D GaAs heterostructure
	\cite{keser_geometric_2021}  to date indicate deviations from predictions of RPA and improved RPA Hubbard theory (See also Ref.~\cite{wang_hydrodynamic_2022,alekseev_viscosity_2020}). To study this deviation from theory and fully characterize and control the electron fluid, one needs an effective and precise method to measure viscosity.
	
	To explain the limitations of the existing viscosity  measurements in Fig.~\ref{AD}  we copy two figures from  Ref.~\cite{keser_geometric_2021}. Simply, the red line in Fig.~\ref{AD}a
	approximately represents the
	ohmic resistance (scattering from impurities and phonons) and the blue line
	represents the total (ohmic plus viscous) resistance.
	At low temperatures, $T \lesssim 20-25$K the difference between the ohmic and total
	resistance is significant, but this is a manifestation of the transition from the
	ballistic to the hydrodynamic regime. The transition itself is an interesting
	problem, see Ref.~\cite{bandurin_fluidity_2018}, but it is not directly relevant to the
	hydrodynamic regime, which is  realized  at $T \gtrsim 20-25$K in this sample.
	In the hydrodynamic regime in this sample, the electron-electron scattering length is
	$l_{ee}\sim 100$ nm and the impurity-phonon scattering length is
	$l_{imp/ph}\gtrsim 5\:\mu$m. While the hydrodynamic regime is realised as $l_{ee} \ll l_{imp/ph}$, according to Fig.~\ref{AD}a,
	the relative contribution of the viscous dissipation is approximately 10\% of
	the ohmic contribution, a relatively small contribution.
	The suppression is easy to understand: the ohmic dissipation exists throughout the sample, while the viscous dissipation comes only from the region near the boundaries~\cite{landau_fluid_2013,moessner_pulsating_2018,levchenko_transport_2020} that contain non-zero
	vorticity. This vorticity is generated  by obstacles,~\cite{gusev_stokes_2020,lucas_stokes_2017} inhomogeneities in the geometry~\cite{moessner_boundary_2019, keser_geometric_2021} or rough channel walls.~\cite{sulpizio_visualizing_2019} Boundary layers have width  $\sim D= \sqrt{l_{ee} l_{imp/ph}/4}$  and constitute a small fraction of the total volume,
	as
	shown in Fig.~\ref{AD}b.  While this example is specific to the experiment
	in Ref.~\cite{keser_geometric_2021}, to the best of our knowledge the issue of the low vorticity 
	is relevant to all existing measurements in graphene and in semiconductors.

	The idea of the present work is to create an oscillatory flow pattern, or ``zigzag''  flow, schematically
	shown in
	%%%%%%%%%%%%%%%%%%%%%%%%%%%%%%%%%%%%%%%%%%%%%%%%%
	\begin{figure}[ht!]
		\includegraphics[width=0.45\textwidth]{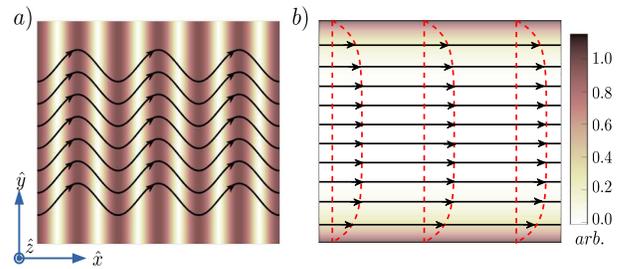}
		\caption{{Velocity field $\mathbf{v}$ (arrows) and streamlines (solid black lines) in a) the ``zigzag'' flow versus b) in ordinary flow inside a no-slip channel, where tangential velocity is zero at the boundaries. In (b) the velocity is purely horizontal and is uniform in the bulk, but the magnitude decays to zero near the no-slip boundaries, forming a non-uniform profile (red dashed lines). Therefore, the strength of vorticity $|\hat{z}\cdot \nabla\times \mathbf{v}|$ (normalized and shown by pink shade) is distributed throughout the bulk in (a), whereas in (b)  it is exponentially small except in the boundary layers.}
		}
		\label{zigzag}
	\end{figure}
	%%%%%%%%%%%%%%%%%
	Fig.~\ref{zigzag}. In the flow through  a channel with non-uniform or rough boundaries, a profile develops that is schematically illustrated by the dashed red lines shown in Fig.~\ref{zigzag}. In such a conventional flow, vorticity is limited to regions near the boundaries of the channel. On the contrary, the solid black oscillatory streamlines in Fig.~\ref{zigzag} carry a uniform vorticity throughout the bulk area, and hence the
	viscous dissipation can match or exceed the ohmic dissipation. 
	
	Periodic geometric modulation or obstacles, create boundary layers near geometric features which decay into the bulk~\cite{gusev_stokes_2020,keser_geometric_2021}
	(see for example red regions in Fig.~\ref{AD}b). Therefore, the ``zigzag'' flow in the bulk cannot be established through geometric modulation. However, this non-decaying ``zigzag'' flow profile may be generated using micromagnets
	that produce a spatially non-uniform magnetic field of amplitude in the range $20$-$30$ mT, that permeates the bulk.
	The spatial period of the magnetic field must be of the order of a few $\mu$m, similar to the characteristic viscous length scale. 
	In principle, the idea of the oscillatory flow is equally applicable to the
	two-dimensional electron gas (2DEG) in  both semiconductor quantum wells and graphene.
	However, boundary conditions that reconcile the oscillatory flow with flat
	boundaries are significantly different for these two cases. In this
	paper, we consider the semiconductor 2DEG in a straight channel with
	no-stress boundary condition.

	The structure of the paper is as follows.
	In  Section~\ref{sec:infinite_channel}
	we  solve the Stokes equation exactly for an infinitely wide 2D flow
	channel with a sinusoidally modulated perpendicular magnetic field.
	The direction of modulation coincides with the direction of the flow.
	This allows us to calculate the viscous and the ohmic dissipation rates
	and to show that the ratio of viscous to ohmic dissipation can be tuned to become large.
	Any realistic flow channel has a finite width, therefore in  Section~\ref{sec:boundary_layer} we
	study the boundary layer in a straight channel using the no-stress
	boundary condition. Through this solution we find the flow inside the boundary layer.
	Since the boundary layer problem is the most mathematically delicate issue,
	we solve the problem by two different methods, (i) perturbation theory,
	(ii) exact numerical solution.
	The central result of Section~\ref{sec:boundary_layer} is the width of the boundary
	layer.  This proves that the ``zigzag'' flow can be realized in  a
	sufficiently wide straight channel and allows the determination of an appropriate channel width.
	In Section~\ref{sec:magnets} we calculate the magnetic field of periodic ferromagnetic stripes
	and hence determine parameters of micromagnets necessary for the experimental
	set up.
	Section ~\ref{sec:concl} presents our conclusions.\\
	
	\section{Flow in infinitely wide and infinitely long 2D channel}
	\label{sec:infinite_channel}
	We consider low currents in the 2DEG, corresponding to low Reynolds number flow. Therefore, we will neglect
	the Navier term,  $\mathbf{v}\cdot\nabla \mathbf{v}$, in the Navier-Stokes equation as it is quadratic in velocity. The flow is also incompressible as the charge carrier density in the 2DEG is controlled via the top gate voltage. Furthermore, we assume steady flow has been established in the channel.
	The resulting stationary Stokes  equation for a two-dimensional incompressible
	electron fluid in a magnetic field reads~ \cite{landau_fluid_2013}
	\begin{align}
		\label{NS}
		\mathbf{v}/\tau 
		-\nu \nabla^2 \mathbf{v}  &= -\nabla \Phi/m^* + q/m^* \mathbf{v}\times \mathbf{B}, \nonumber\\
		\nabla \cdot \mathbf{v} &=0.
	\end{align}
	Here $\mathbf{v}$ is the fluid velocity, $\nu$ is the kinematic viscosity,
	$\tau$ is the relaxation time related to ohmic processes (scattering from
	impurities and phonons), $\Phi$ is the electro-chemical  potential, $q$ is
	the charge of the fluid particle, $m^*$ is the effective hydrodynamic
	mass and $\mathbf{B}$ is the magnetic field. The magnetic field perpendicular to the plane of
	the 2DEG is periodically modulated along the flow as
	\begin{align}
		\label{bx}
		B=B_z&=B_0b(x),\nonumber\\
		b(x)&=\sin(gx),\nonumber\\
		g&=\frac{2\pi}{a}.
	\end{align}
	For electrons in GaAs, considered in the present paper, the mass $m^*$ is
	just the band structure mass, $m^*\approx 0.067m_e$.
	Using the no-stress boundary condition method developed in Ref.\cite{keser_geometric_2021}
	the relaxation time $\tau$ can be measured independently
	of hydrodynamics. $\tau$  is a known phenomenological function of
	temperature. It is convenient to relate the viscosity and the relaxation
	time to the electron-electron scattering length $l_{ee}$ and to the
	electron mean free path $l_{mfp}$.
	\begin{align}
		\label{nut}
		\nu&=\frac{1}{4}v_Fl_{ee},\nonumber\\
		\tau&=\frac{l_{mfp}}{v_F}.
	\end{align}
	The relations~\eqref{nut} are well known at low temperatures,
	$T \ll \epsilon_F$, where $\epsilon_F$ is the Fermi energy.
	We extend~\eqref{nut} to higher temperatures, $T \sim \epsilon_F$, using these relations as definitions of the corresponding lengths, where we use the Fermi velocity $v_F$  defined at zero temperature,
	\begin{eqnarray}
		\label{vf}
		v_F=\frac{p_F}{m^*}=\frac{\sqrt{2\pi n}}{m^*}.
	\end{eqnarray}
	Here $n$ is the number density of the 2DEG and $p_F$ is the Fermi momentum.

	The viscosity depends on magnetic field~\cite{alekseev_negative_2016} as,
	\begin{eqnarray}
		\label{bstar}
		\nu \propto \frac{1}{1+(B/B^*)^2}.
	\end{eqnarray}
	Here, $B^*$ is the characteristic magnetic field of the material, related to the electron-electron scattering length by $B^*  = {p_F}/{(2|e|l_{ee})}$.~\cite{alekseev_negative_2016} Although this suppression of viscosity by the magnetic field can be incorporated in our analysis, we consider 
	very weak fields, $B\sim B^*/10$, so we disregard the $B$-dependence of $\nu$.
	
	To solve Eq.~\eqref{NS} it is convenient to use the stream function, $\psi$,
	defined as
	\begin{align}
		\label{str}
		v_x&=\partial_y\psi,\nonumber\\
		v_y&=-\partial_x\psi,
	\end{align}
	which ensures that $\nabla \cdot \mathbf{v}=0$. Acting on Eq~\eqref{NS} with the curl operator $\nabla \times $, and
	using the definition Eq.~\eqref{str}, we obtain
	\begin{subequations}
		\begin{align}
			\label{NS1}
			- \nabla ^2 \psi + D^2 \nabla ^4 \psi &= \omega \tau (\partial_x b)
			\partial_y \psi, \\
			\omega=\frac{|q|B_0}{m^*},&\quad
			D=\sqrt{\nu\tau}=\frac{1}{2}\sqrt{l_{ee}l_{mfp}}.
		\end{align}
	\end{subequations}
	
	We first consider an applied current density
	$\mathbf{J} = q n v_0 \hat{x}$ that passes through a very large system and find
	\begin{equation}
		\label{pinf}
		\psi_\infty = v_0 \left( y + \frac{\omega \tau}{g (1 + D^2 g^2)} \cos (gx)\right),
	\end{equation}
	where the subscript $\infty$ indicates that this is the solution when system width and length tend to infinity.
	Indeed, the stream function in Eq.~\ref{pinf} generates the zigzag flow shown in
	Fig.~\ref{zigzag}. Note that if the magnetic field has many Fourier components,
	$b(x)=\sum_{n=1}^{\infty}[\alpha_n\sin(ngx)+\beta_n\cos(ngx)]$,
	the generalization of Eq. ~\eqref{pinf} is straightforward.
	
	Instead of determining the relationship between electrochemical potential and current from  Eq.~\eqref{NS}, we can infer the resistance in the system by accounting for the total power dissipation. The terms $m^* \mathbf{v}/\tau$ and $-m^* \nu\nabla^2 \mathbf{v}$ in Eq.~\eqref{NS} are respectively the ohmic drag and viscous friction forces in the system. Therefore, the force velocity product gives the corresponding dissipation rates which follow from Eqs.~\eqref{NS} and~\eqref{pinf} as
	\begin{eqnarray}
		\label{diss}
		\frac{\dot{E}_\Omega}{LW} =\frac{nm^*}{LW \tau}\int d^2r v^2=\frac{nm^*v_0^2 }{\tau}\ \left(1+  \frac{\omega^2 \tau^2}{2(1+D^2 g^2)^2}
		\right) ,\nonumber \\
		\frac{\dot{E}_\nu}{LW} =\frac{-nm^*\nu}{LW}\int d^2r{\bm v}\cdot \nabla^2{\bm v}=
		\frac{nm^*v_0^2 }{2\tau}\ \left( \frac{gD \omega \tau}{1+D^2 g^2}\right)^2.
	\end{eqnarray}
	Here $L$ and $W$ are the length and the width of the channel respectively.
	
	We may tune free parameters to maximize the ratio $\dot{E}_\nu/  \dot{E}_\Omega$ such that viscous transport is not overshadowed by ohmic effects.
	It is easy to check that at a fixed value of $\omega\tau$ the ratio is maximum 
	at the following value of $gD$
	\begin{eqnarray}
		\label{ratio1}
		&&t=\omega\tau,\nonumber\\
		&& gD= [1+t^2/2]^{1/4},\nonumber\\
		&&R=\frac{\dot{E}_\nu}{\dot{E}_\Omega}=\frac{t^2}{4(1+\sqrt{1+t^2/2})}.
	\end{eqnarray}
	At large magnetic field the ratio grows linearly with amplitude of the magnetic field,
	$R\approx \omega\tau/(2\sqrt{2})$. However, to avoid the influence of the magnetic
	field on the viscosity, Eq.~\eqref{bstar}, we need $B_0 \ll B^*$.
	This means that the optimization of parameters depends on the value of the relaxation
	time $\tau$ that is related to the  mobility of the 2DEG.
	
	We make realistic estimates for the effect of periodic magnetic field on the  GaAs system from Ref.~\cite{keser_geometric_2021}. In order to pick the optimum value for the period of the magnetic superlattice, we first determine $\omega\tau$ from the experimentally measured parameters of the 2DEG. At $T=30-40$K  and density $n\sim few\times 10^{11}\:\text{cm}^{-2}$ has 
	$B^*\approx 200-250$mT,  $\tau \approx 0.3\times 10^{-10}$s and the corresponding
	mean free path $l_{mfp}\approx 7\mu$m.
	If we take $B_0=30$ mT the $B^*$ correction in Eq.~\eqref{bstar} to $\nu$ at the peak magnetic field strength is approximately 2\%
	and can be safely neglected. With these values of  $B_0$ and $\tau$ we find
	$\omega\tau=2.4$ and hence according to Eq.~\eqref{ratio1} the ratio is $R\approx 0.5$.
	Hence, we achieve almost an order of magnitude enhancement of the viscous dissipation
	compared to Fig.~\ref{AD}a. The value of the characteristic length $D$ defined in
	Eq.~\eqref{NS1} is $D\approx 0.6\mu$m. Finally, according to Eq.~\eqref{ratio1}
	the optimal period of the magnetic superlattice is $a = 2\pi/g \approx 2.6\:\mu$m.\\
	
	\section{Boundary layer for a straight channel with no-stress boundary condition}
	\label{sec:boundary_layer}
	
	In the previous section we considered an infinite fluid domain.
	Of course, any physical channel must have some finite width $W$. Here we consider a straight channel with
	no-stress boundary conditions engineered experimentally in Ref.~\cite{keser_geometric_2021}.
	To support the message of the previous section, we seek to prove that the boundary layer that
	reconciles the oscillatory bulk flow with the straight flow near the boundary is of a finite
	width.  In other words, the width of the boundary layer must remain finite as the amplitude of magnetic field is increased.  
	%The central message of the present section is that indepenent of the amplitude
	%of the field (the amplitude of the flow zizag modulation) the width of the layer is about
	%the characteristic length $D$, see Eq.(~\ref{NS1}), so the bulk approximation is justified if $D \ll W$.
	Because of the importance of this assertion, we check it by two different methods, (i) analytic
	solution using perturbation theory in magnetic field, and (ii) exact numerical solution.
	
	\subsection{Perturbation theory for the boundary layer flow}
	\label{sec:pert_boundary_layer}
\begin{figure}[ht!]
		\includegraphics[width=0.23\textwidth]{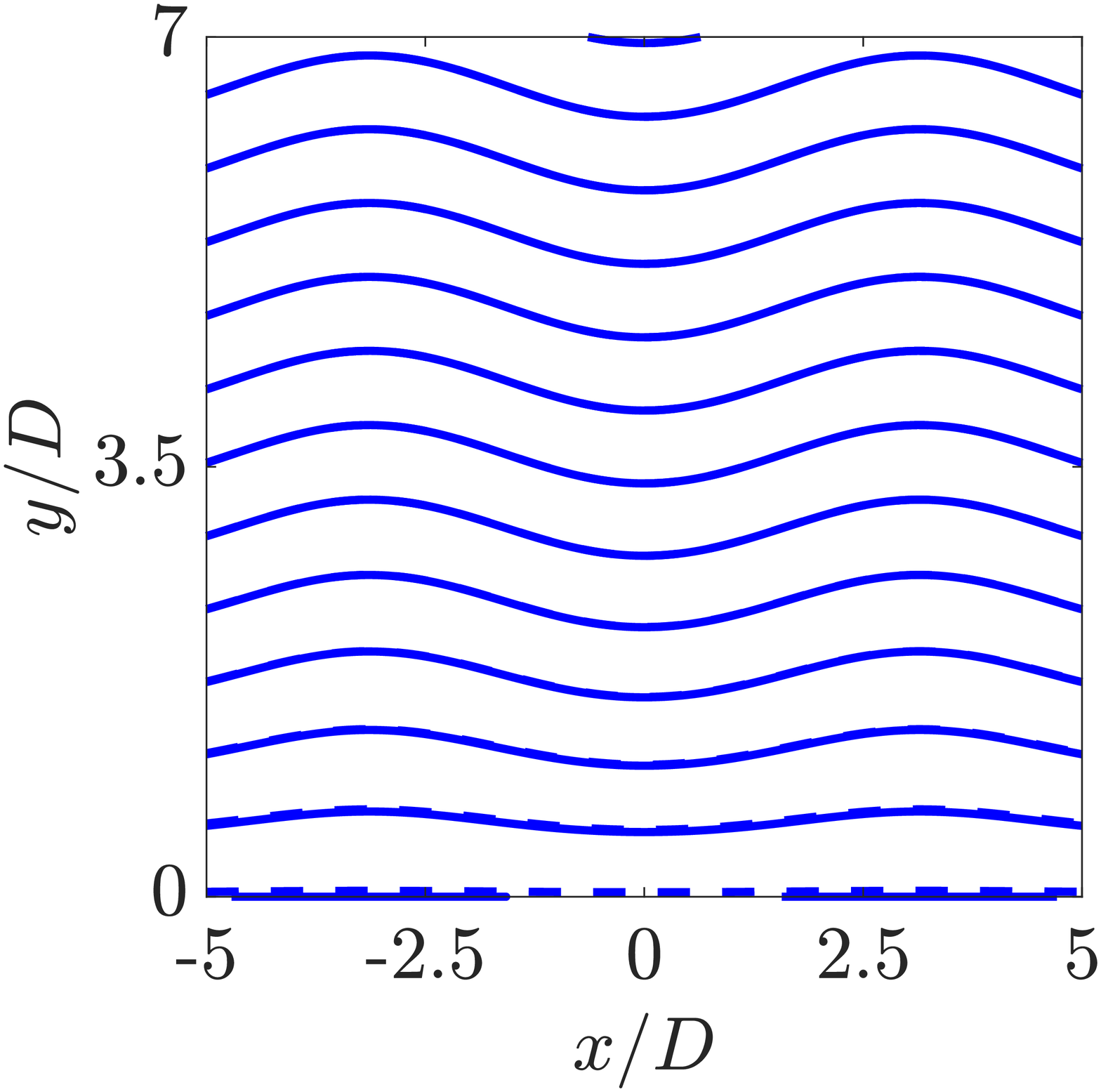}
		\includegraphics[width=0.23\textwidth]{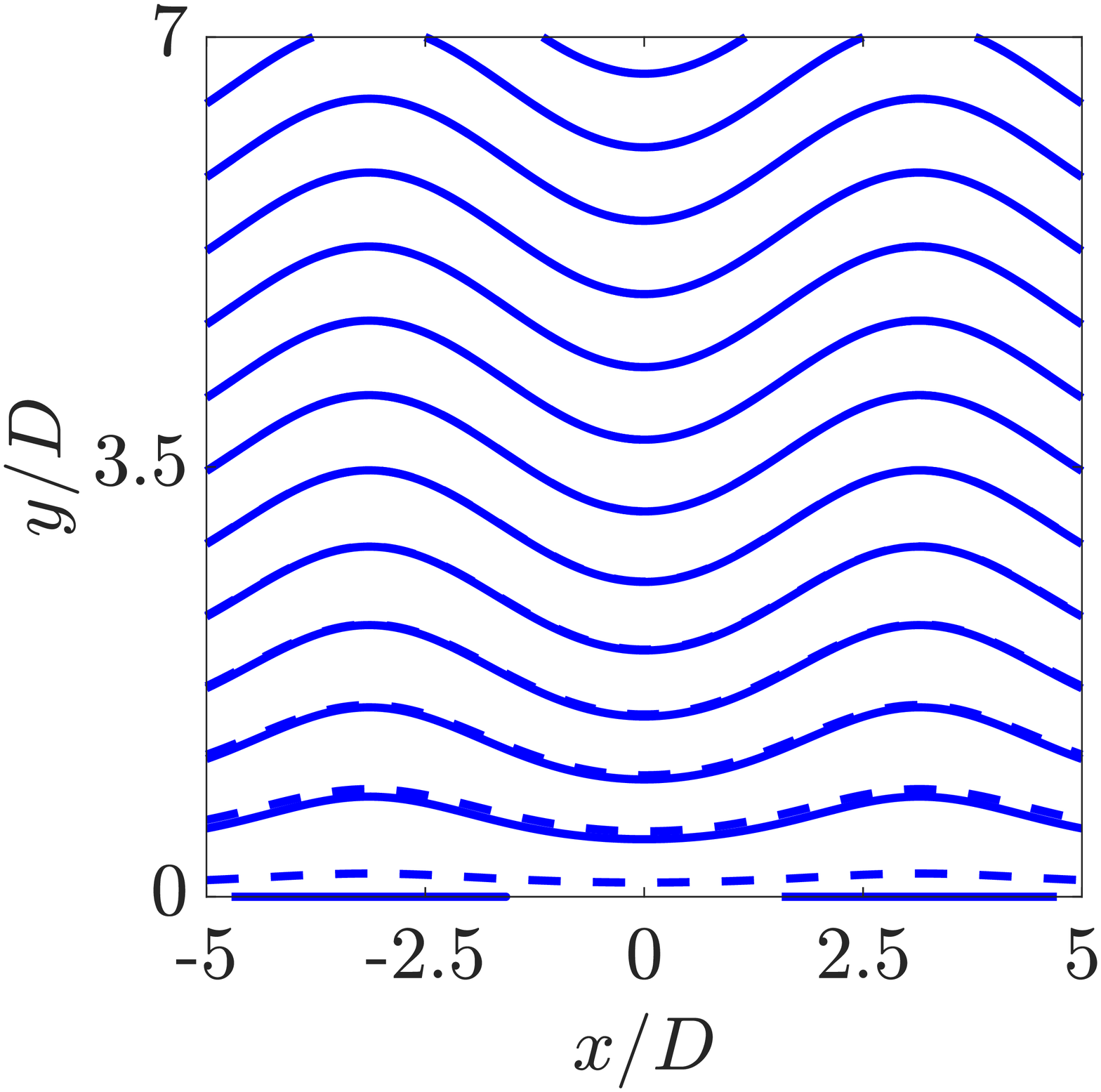}
		\caption{Streamlines for $gD=1$ and $\omega\tau=0.5$ (left) and $\omega\tau=1$ (right).
			Solid lines correspond to the 1st order perturbation theory and dashed lines correspond to the
			2nd order perturbation theory. In both cases the dashed lines almost exactly coincide with the solid lines for $y>D$. 
		}
		\label{pt}
	\end{figure}

	For the purpose of deriving an analytic expression for the boundary, we consider the boundary region
	as a semi-infinite channel with domain  $y>0$ and a perfect-slip boundary at $y=0$.
	%  The expression derived thus holds true for a channel with width much greater
	%than the effective width of the boundary layer.
	Together with the Stokes equation, Eq.~\eqref{NS1},  the stream function must satisfy the no-penetration and perfect-slip, or no-stress, boundary conditions as follows, respectively,
	\begin{equation}
		\label{bc}
		v_y|_{y=0}=-\partial_x\psi|_{y=0} = 0, \; \partial_yv_x|_{y=0}=\partial_y^2\psi|_{y=0} = 0.
	\end{equation}
	The former, no-penetration condition ensures no fluid can enter/exist through the boundary. Whereas the latter no-stress condition ensures that the fluid experiences no friction from the boundary, in other words the perfect-slip condition is realized.  Furthermore, the stream function must decay to the bulk solution~\eqref{pinf} far from the boundary.
	We look for a solution as an expansion in the powers of $\omega\tau \ll 1$,
	$\psi = \sum_{n=0}^{\infty} (\omega\tau)^n \psi^{(n)}$. Straightforward substitution in Eqs.~\eqref{NS1},and ~\eqref{bc} gives the following 1st order solution
	\begin{eqnarray}
		\label{1st}
		\psi &=& v_0\left\{ \left(y + \frac{\omega \tau}{g(1+g^2D^2)}\cos(gx)\right)\right.\\ 
		&+& \left.\frac{\omega \tau}{g}\left(\frac{g^2D^2}{1+g^2D^2}e^{-\sqrt{g^2D^2 + 1}y/D}-e^{-g y}\right)
		\cos(gx) \right\}.  \nonumber
	\end{eqnarray}
	Due to the optimization condition~\eqref{ratio1} the wave vector is $g > 1/D$, therefore the boundary
	layer part of the solution~\eqref{1st} decays faster than the characteristic length $D$.
	In Appendix ~\ref{app_1} we present the expansion of the solution to $\psi$ to the second order in $\omega\tau$ to demonstrate that the effective width of higher order contributions to the boundary decreases with increasing order.
	The corrections to $\psi$ to second order in $\omega\tau$ part has boundary layer terms that exponentially decay with $y$ at the same rate as that in Eq.~\eqref{1st} or faster. Fig.~\ref{pt} presents plots of streamlines obtained in the 1st and the 2nd order perturbation theory at $\omega\tau=0.5,1$ and $gD=1$.
	%%%%%%%%%%%%%%%%%%%%%%%%%%%%%%%%%%%%%%%%%%%%%%%%%
	%%%%%%%%%%%%%%%%%
	The difference between the solid and the dashed lines in Fig.~\ref{pt} characterize the convergence
	of the perturbation theory. While at $\omega\tau=0.5$ there is practically no visible difference between the  first and second order perturbation theory, the difference is visible when $\omega\tau=1$. In both cases the boundary layer has almost entirely decayed for $y> 2D$, with the flow indistinguishable from the bulk solution given by Eq.~\eqref{pinf}.
	
	The perturbation theory at $\omega\tau=1$ does not satisfy the small $\omega\tau$ limit and it is hardly  possible to justify the perturbation theory at $\omega\tau\approx 3$ corresponding to the optimal device. Therefore, in the following subsection we present the numerical solution valid at arbitrary
	$\omega\tau$.
	
	\subsection{Exact numerical solution for the boundary layer flow}
	\label{sec:exact}

	\begin{figure*}[th!]
		\includegraphics[width=0.28\textwidth]{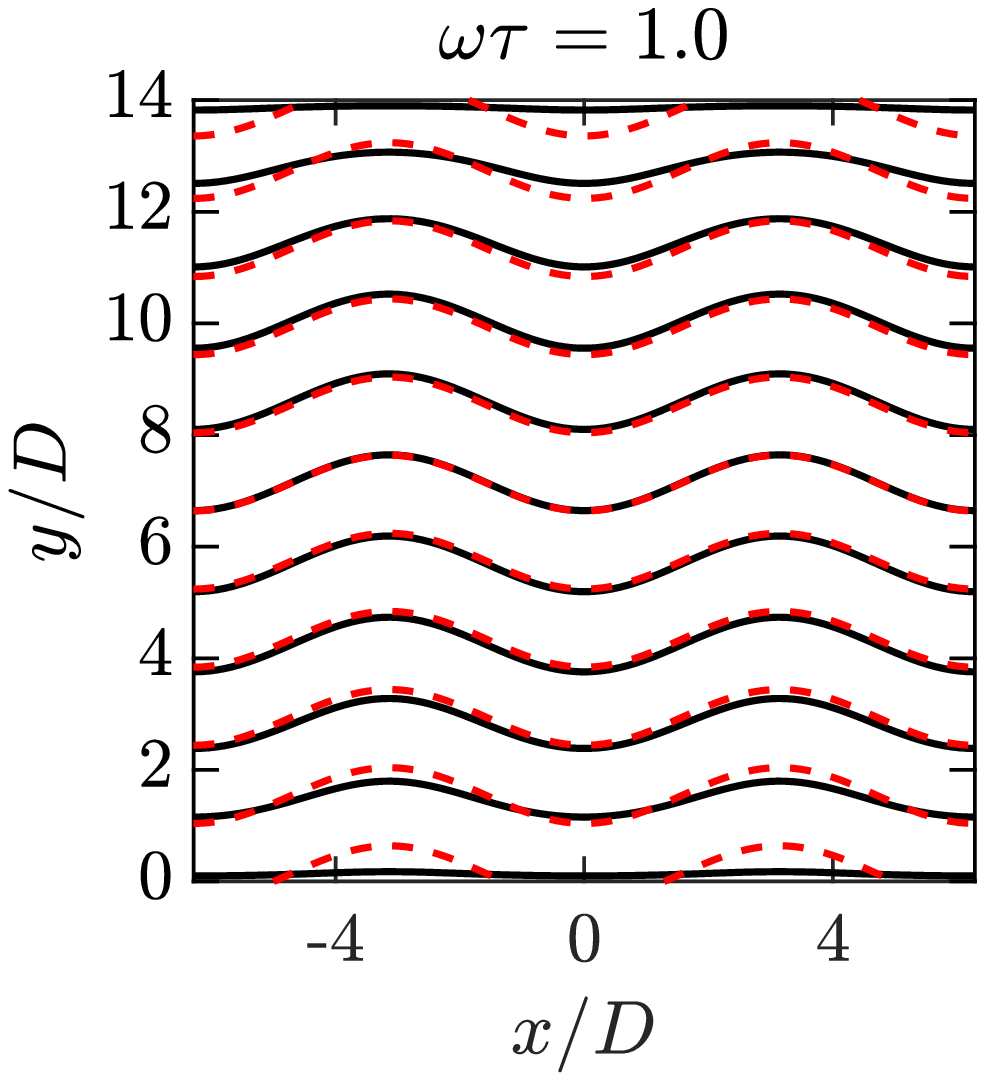}
		\hspace{0.7cm}
		\includegraphics[width=0.28\textwidth]{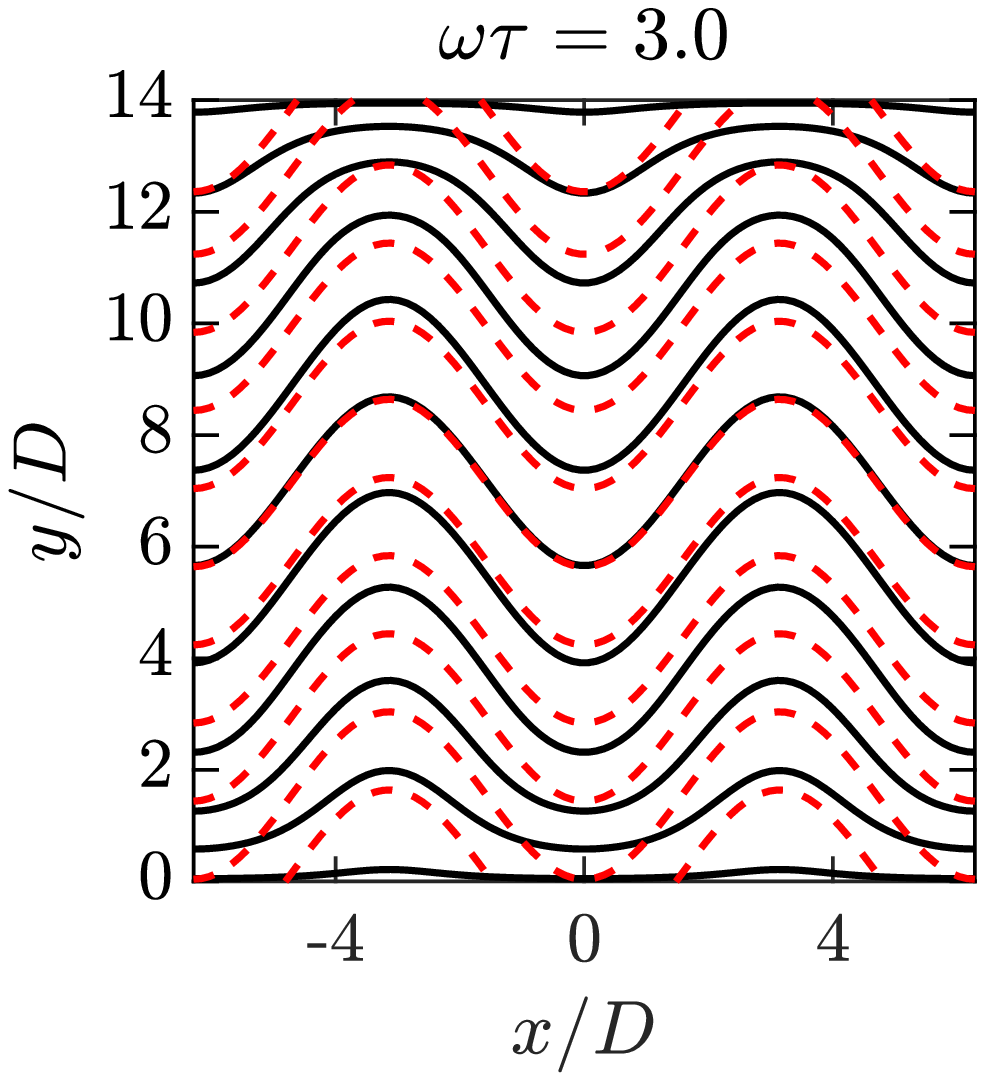}
		\hspace{0.7cm}
		\includegraphics[width=0.28\textwidth]{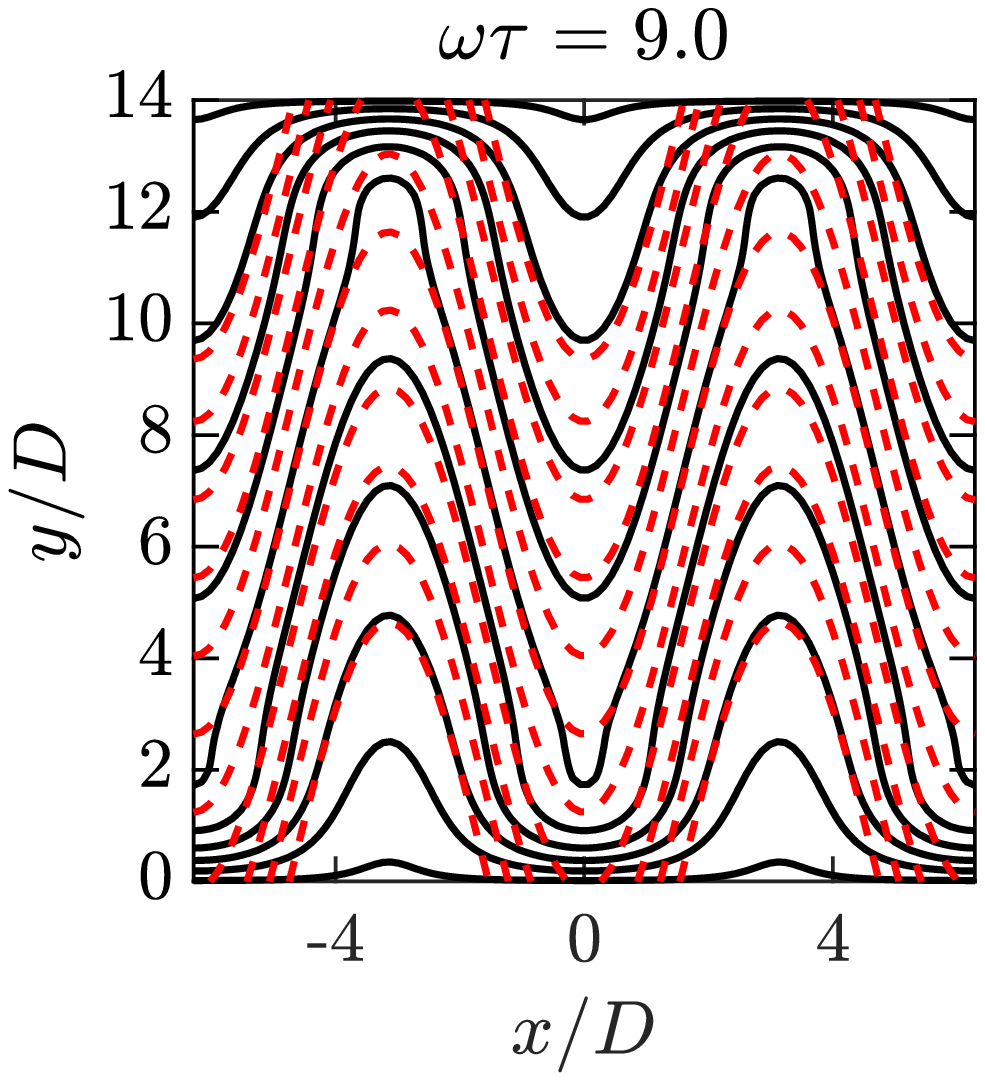}
		\caption{Streamlines in a channel with no-stress boundaries with width $W = 14 D$. The parameters of the magnetic field are chosen as $gD=1$,
			$\omega\tau=1, 3, 9$.
			Solid lines correspond to the numerically exact solution described in
			Sec.~\ref{sec:exact} and dashed lines correspond to the solution given by Eq.~\eqref{pinf} in the
			infinite device.
		}
		\label{fig:exact}
	\end{figure*}

	In this section, we reduce the Stokes equation, Eq.~\eqref{NS1}, to an infinite
	sequence of linear algebraic equations. We then numerically solve  these equations by truncating them at a suitable order.  We consider an infinitely long channel of width $W$ and use length units where $D =\sqrt{\nu \tau}= 1$.
	As the magnetic field is periodic in the $x$-direction so is the stream function, $\psi$, which can be presented as the following Fourier series,
	\begin{eqnarray}
		\psi = v_0 y + v_0 W\sum_{m=-\infty}^{\infty}e^{i g m x } G_m(y)\ .	
	\end{eqnarray}
	Here $G_m(y)$ are unknown functions to be determined.
	The first term $v_0 y$ corresponds to the uniform velocity $v_x=v_0\neq 0,\:v_y = 0$.
	
	At $y=0$ and, $y=W$ the stream function must satisfy the no-penetration 
	and no-stress boundary conditions, previously given for $y=0$ in Eq.~\eqref{bc}.
	The no-penetration condition implies that $G_m(0)=G_m(W)=0$. Hence, the functions
	$G_m(y)$ can be expanded in a Fourier series in
	$\sin\left(n y\pi/W\right)$ as
	\begin{eqnarray}
		\psi/v_0 = y +  W\sum_{n=1}^{\infty}\sum_{m=-\infty}^{\infty} f_{m,n} e^{i g m x } \sin\left(\frac{n y\pi}{W}\right),	
	\end{eqnarray}
	where $f_{m,n}$ are complex coefficients.
	Note that, this form is chosen so that the no-stress condition is automatically
	satisfied. Furthermore, the sinusoidal terms do not contribute to the average current through the channel.
	The Navier-Stoke's equation, Eq.~\ref{NS1}, expressed in terms of the coefficients
	$f_{m,n}$ reads
	\begin{multline}
		\label{NSalg}
		\sum_{n=1} D_{m,n} f_{m,n}  \sin(n y\pi/W) = \frac{\alpha}{2}\bigg(\delta_{m,1} + \delta_{m,-1} \\+ \sum_{n=1} \left[f_{m+1,n} + f_{m-1,n}\right]  n\pi\cos(n y\pi/W) \bigg), 
	\end{multline}
	where $\alpha = -\omega\tau g/W$ and
	\begin{equation}
		D_{m,n} = - \left[\left(m g \right)^2 + \left(\frac{\pi n}{W} \right)^2 \right]\left[\left( m g  \right)^2 + \left(\frac{\pi n}{W}\right)^2  +1\right].
	\end{equation}
	Eq.~\eqref{NSalg} is defined only at $0<y<W$. Therefore, one cannot naively
	equate the Fourier components. Instead we must integrate over $y$
	using the following identities
	\begin{eqnarray}
		&&	\int_0^{1}dy\: \sin(n y \pi) \sin( l y\pi) =
		\frac{1}{2}\delta_{n,l},\nonumber\\
		&&  \int_0^{1}dy\: \cos(n y \pi) \cos(l y\pi)=
		\frac{1}{2}\delta_{n,l},\nonumber\\
		&&\int_0^{1}dy\: \sin(n y \pi) \cos(l y\pi) =F(n,l),\nonumber\\
		&&F(n,l)= \left\{  \begin{array}{ccc}
			0 &  ,  &n+l=even\\
			\frac{2n}{\pi(n^2-l^2)}& , & n+l=odd.
		\end{array}
		\right.
	\end{eqnarray}
	Hence, Eq.(~\ref{NSalg}) is reduced to the following system of linear
	algebraic equations for the coefficients $f_{m,l}$:
	%\begin{eqnarray}
	%  \label{NSalg1}
	% D_{m,l} f_{m,l} &&= {\alpha}\bigg(
	%  (\delta_{m,1} + \delta_{m,-1})F(l,0)\\
	%&&+ \sum_{n=1}^{\infty}
	% \left[f_{m+1,n} + f_{m-1,n}\right]  n\pi F(l,n) \bigg),\nonumber \\
	%  \sum_{n=1}^{\infty}D_{m,n} f_{m,n}F(n,l) && = \frac{\alpha}{4}\bigg(\left[f_{m+1,l} + f_{m-1,l}\right]  (l\pi)   \\
	%&&+ \left[f_{m+1,n} + f_{m-1,n}\right] \delta_{l,0}\bigg).\nonumber
	%\end{eqnarray}
	
	\begin{subequations}
		\label{NSalg1}
		\begin{align}
			\begin{split}
				D_{m,l} f_{m,l} = {\alpha}&\bigg(\sum_{n=1}^{\infty}
				\left[f_{m+1,n} + f_{m-1,n}\right]  n\pi F(l,n)\\
				&+  (\delta_{m,1} + \delta_{m,-1})F(l,0) \bigg),\: l>0,
			\end{split}\\
			\begin{split}
				\sum_{n=1}^{\infty}D_{m,n} f_{m,n}&F(n,l)  = \frac{\alpha}{4}\bigg(\left[f_{m+1,l} + f_{m-1,l}\right]  l\pi  \\
				&+ \left[f_{m+1,n} + f_{m-1,n}\right] \delta_{l,0}\bigg),\: l\geq 0.
			\end{split}
		\end{align}
	\end{subequations}
	
	%Obviously $f_{m,l}=f_{-m,l}$, therefore  Eqs.(~\ref{NSalg1}) have to be solved
	%for $0\leq m < \infty$ and $1 \leq l < \infty$. 
	We truncate beyond $|m|,l,n > 50$. With a channel width $W =14$ the truncation
	implies spatial uncertainty $\sim 14/50 \approx 0.3$, which is accurate
	enough to describe the boundary layer.
	
	Plots of the streamlines obtained
	from this numerical solution for $gD=1$ and $\omega\tau=1,3,9$ are presented in Fig.~\ref{fig:exact}.

	\begin{figure}[ht!]
		\includegraphics[width=0.4\textwidth]{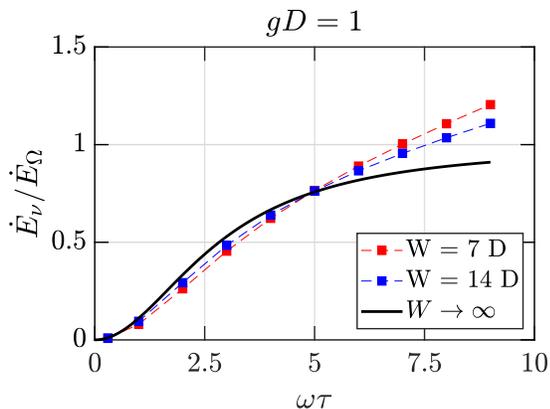}
		\caption{The ratio of viscous to ohmic dissipation at $gD =1$. For the infinite system (solid black line), this ratio is calculated from Eq.~\eqref{diss}. For a system with finite width, $W= 7D$ (red dots) and $W=14 D$ (blue dots), it is calculated from the numerical solution in Sec.~\ref{sec:exact}. Finite width (boundary layer) effects are negligible when $W\gg 2D$ below $\omega\tau \lesssim 6$.
		}
		\label{fig:rat_compare}
	\end{figure}
	
	The exact solution demonstrates that the width of the boundary layer
	increases with increasing magnetic field. For the most interesting case,
	$\omega\tau \approx 3$ the width of the boundary layer is about $2D$. The fact that the boundary layers are growing in size with $\omega \tau$ is also evident in Fig.~\ref{fig:rat_compare}. We compare the ratio of viscous to ohmic dissipation in an infinite channel, as calculated in Eq.~\eqref{diss}, to a finite width channel. In Fig.~\ref{fig:rat_compare}, the ratio calculated from the exact solution approaches that of an infinite system as the channel width grows, as expected. Of course, for small values of $\omega\tau < 1$, the thin boundary layers, as in Fig.~\ref{fig:exact}, have vanishing stress that reduces viscous dissipation. More interestingly, the finite width corrections become noticeable at large $\omega\tau$, where the boundary layer width $2D$ is comparable to the width $W$. 
	Hence, the width of the channel $W$ must be significantly larger than $2D$ for boundary layer effects to be small. For GaAs with mobility of about $10^6$ corresponding to the device
	from Ref.~\cite{keser_geometric_2021} this implies that $ W \gg 1.2\mu$m.

	%\begin{equation}
	%	\label{linear}
	%\mathbb{E}f^{even}  = \alpha\mathbb{L} f^{odd},\quad \mathbb{R} f^{odd} = \alpha{S} + \alpha\mathbb{K}f^{even}.	
	%\end{equation}
	%where
	%\begin{equation}
	%		f^{even}_{m,l} = f_{m,2l},\quad f^{odd}_{m,l} = f_{m,2l-1},\quad l=1,2... 
	%\end{equation}
	%The vector $S$ and the matrices $\mathbb{E,K,L,R}$ are
	%\begin{subequations}
	%\begin{align}
	%{S}_{m,l} &= \frac{1}{2}[\delta_{m,1} + \delta_{m,-1}]\delta_{l,1},\\
	%	\mathbb{E}_{(m,l),(r,n)} &= \delta_{m,r}\delta_{l,n} D_{m,2l},\\ 	 \mathbb{K}_{(m,l),(r,n)} &=    \frac{(l-1)\pi}{2}\left[\delta_{r,m+1} + \delta_{r,m-1}\right]\delta_{n,l-1},\\
	%	\mathbb{L}_{(m,l),(r,n)} &= 	\frac{4l(2n-1)}{(4l^2 - (2n-1)^2)}\left[\delta_{r,m+1} + \delta_{r,m-1}\right],\\	\mathbb{R}_{(m,l),(r,n)} &=  D_{m,2n-1} 	\frac{2(2n-1)}{\pi((2n-1)^2 - 4(l-1)^2)}\delta_{r,m}.
	%\end{align}
	%\end{subequations}
	%We solve the linear system Eq.\eqref{linear} numerically and plot the flow in Fig.~~\ref{fig:exact} for the perturbative case $\omega\tau=gD=1$,
	%   $W= 14D$, and the ``optimal'' device described in Sec.~~\ref{sec:infinite_channel},
	%$\omega\tau= 2.4$, $gD=1.4$, $W\sim14D$.
	%\begin{figure}
	%\centering
	%\begin{subfigure}[b]{0.5\columnwidth}
	%\centering
	%\includegraphics[width=1\columnwidth]{pe%rturb_compare.eps}
	%		\caption{}
	%		\label{fig:perturb_compare}
	%	\end{subfigure}
%\hspace{0.01\columnwidth}
%	\begin{subfigure}[b]{0.5\columnwidth}
	%		\centering
	%		\includegraphics[width=1\columnwidth]{op%timal_exact.eps}
	%		\caption{}
	%		\label{fig:optimal_exact}
	%	\end{subfigure}
%	\caption{}
%	\label{fig:exact} 
%\end{figure}

\section{Magnetic field of periodic ferromagnetic stripes}
\label{sec:magnets}
The calculations in this section demonstrate that realization of the 
optimal magnetic field profile for studies in the hydrodynamic regime is
realistically achievable.

\begin{figure}[ht!]
	\includegraphics[width=0.5\textwidth]{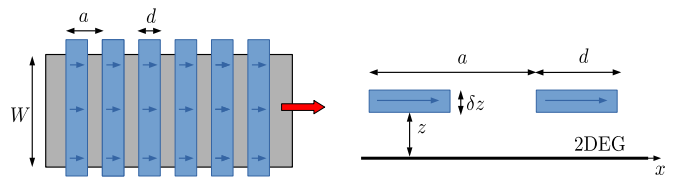}
	%  \hspace{10pt}
	%\includegraphics[width=0.25\textwidth]{BarArraySide.eps}
	\caption{ Left: top view of a periodic array of magnetic bars perpendicular to
		the flow direction shown by the red arrow. The period of the array is $a$ and
		the width of the bar is $d$.  The bars are magnetized parallel to the longitudinal axis of the  channel. The direction of the magnetization is shown by the blue arrows.
		Right: Side view of the device. The vertical distance from the bar magnets to the 2DEG is $z$ and the
		thickness of the rectangular bar is $\delta z$.
	}
	\label{MB}
\end{figure}

We consider a setup consisting of an array of rectangular magnetic bars
perpendicular to the longitudinal axis of the flow channel, as seen below in the left panel of Fig.~\ref{MB}.
%%%%%%%%%%%%%%%%%%%%%%%%%%%%%%%%%%%%%%%%%%%%%%%%%
%%%%%%%%%%%%%%%%%
The bars are magnetized along the longitudinal axis of the channel.
The bar thickness is $\delta z$ and the distance from the base of the bar to the 2DEG is $z$,
shown in the right panel of Fig.~\ref{MB}.
We set the value of the period $a=2.6\mu$m as obtained in
Sec.~\ref{sec:infinite_channel}.
We tune the remaining free parameters $d,z,\delta z$ such that
the z-component of magnetic field in the 2DEG is approximately
$B_0\sin(2\pi x/a)$ with $B_0\approx 30$ mT. This value of $B_0$ limits the magnetic correction to the viscosity to a maximum of approximately 2\% and thus this effect may be safely neglected in the analysis. 
Assuming the magnetic moment is uniform throughout the volume of each bar, and the bars extend in the transverse ($y$) direction wider than the width $W$ of the sample,  we calculate the magnetic field on the plane of 2DEG as a function of longitudinal coordinate $x$, from the Biot-Savart law as
\begin{multline}
	\label{Bfield}
	\frac{B_z(x)}{B_m}\bigg\vert_{2DEG} = -\sum_{n>1} \frac{\sin\left(g n d/2\right)}{ \pi n} \sin(g n x) Z_n,\\Z_n(z,\delta z)= e^{-g n z} (1-e^{-g n \delta z}),
\end{multline}
where $B_{m}$ is the saturation field, and $x$ is measured from the center of one of the magnets. For example, at the edge of the magnet $x=d/2$, the harmonics of the field on 2DEG plane is $\propto \sin^2(g  n d/2)$. Therefore, $d=a/2 = \pi/g$ maximizes the first harmonic and suppresses the second harmonic. We numerically calculate $B$ from Eq~\eqref{Bfield} and further confirm that $d=a/2=1.3\:\mu$m is optimal. In other words, matching the width of the magnet to the spacing between successive magnets creates an effectively sinusoidal magnetic field profile without higher harmonics. 

We consider two magnetic materials, (i) Ni that has saturation
magnetic field 600mT \cite{danan_new_1968} and (ii) NiFe alloy that 
has saturation magnetic field 1500mT. \cite{tumanski_magnetic_2011}
We consider two values of bar thickness, $\delta z =0.3\:\mu$m and
$\delta z =0.2\:\mu$m. After fixing these values, the only free parameter is the distance to 2DEG $z$, which we tune.

%%%%%%%%%%%%%%%%%%%%%%%%%%%%%%%%%%%%%%%%%%%%%%%%%
\begin{figure}[ht!]
	\includegraphics[width=0.23\textwidth]{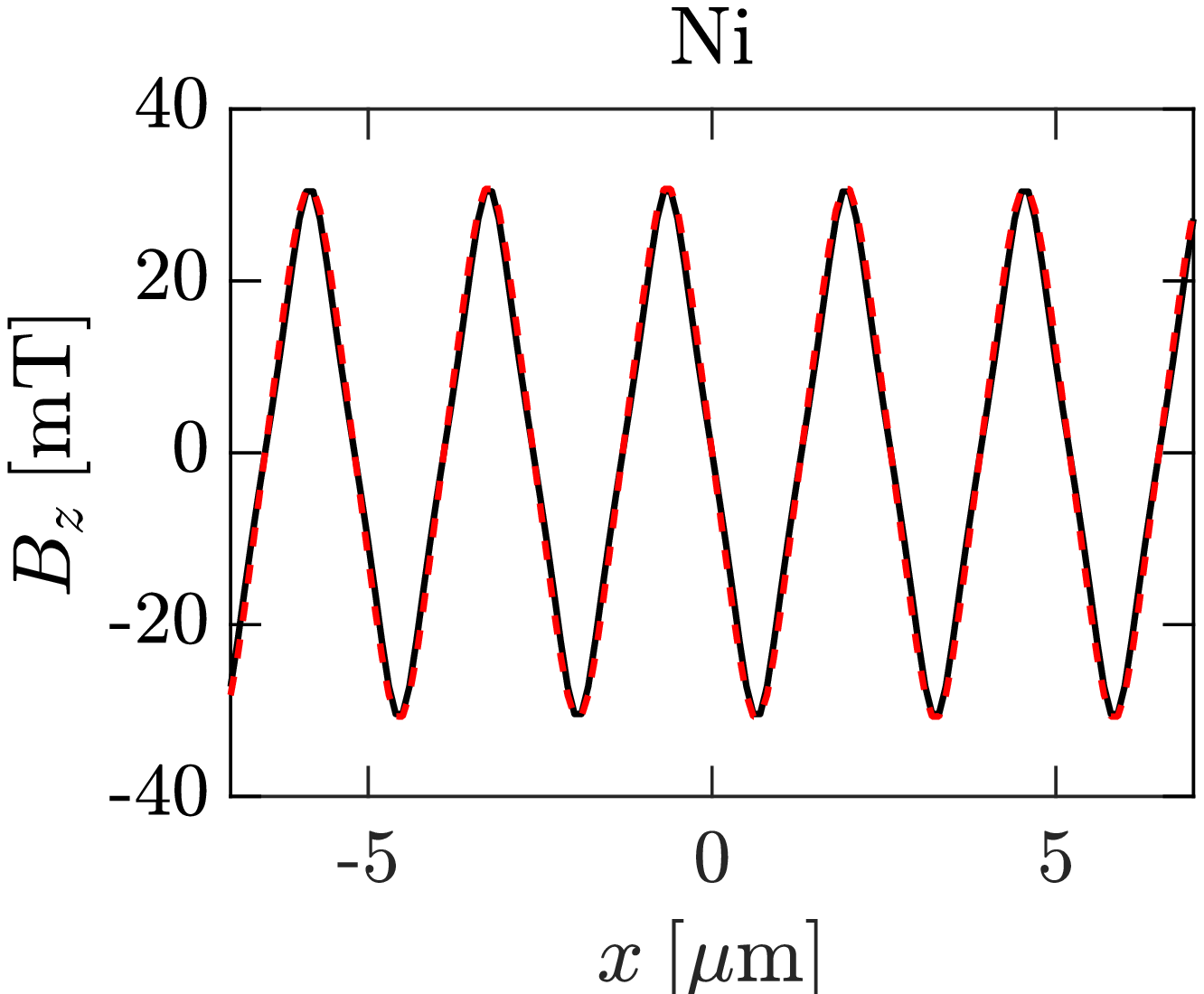}
	\includegraphics[width=0.23\textwidth]{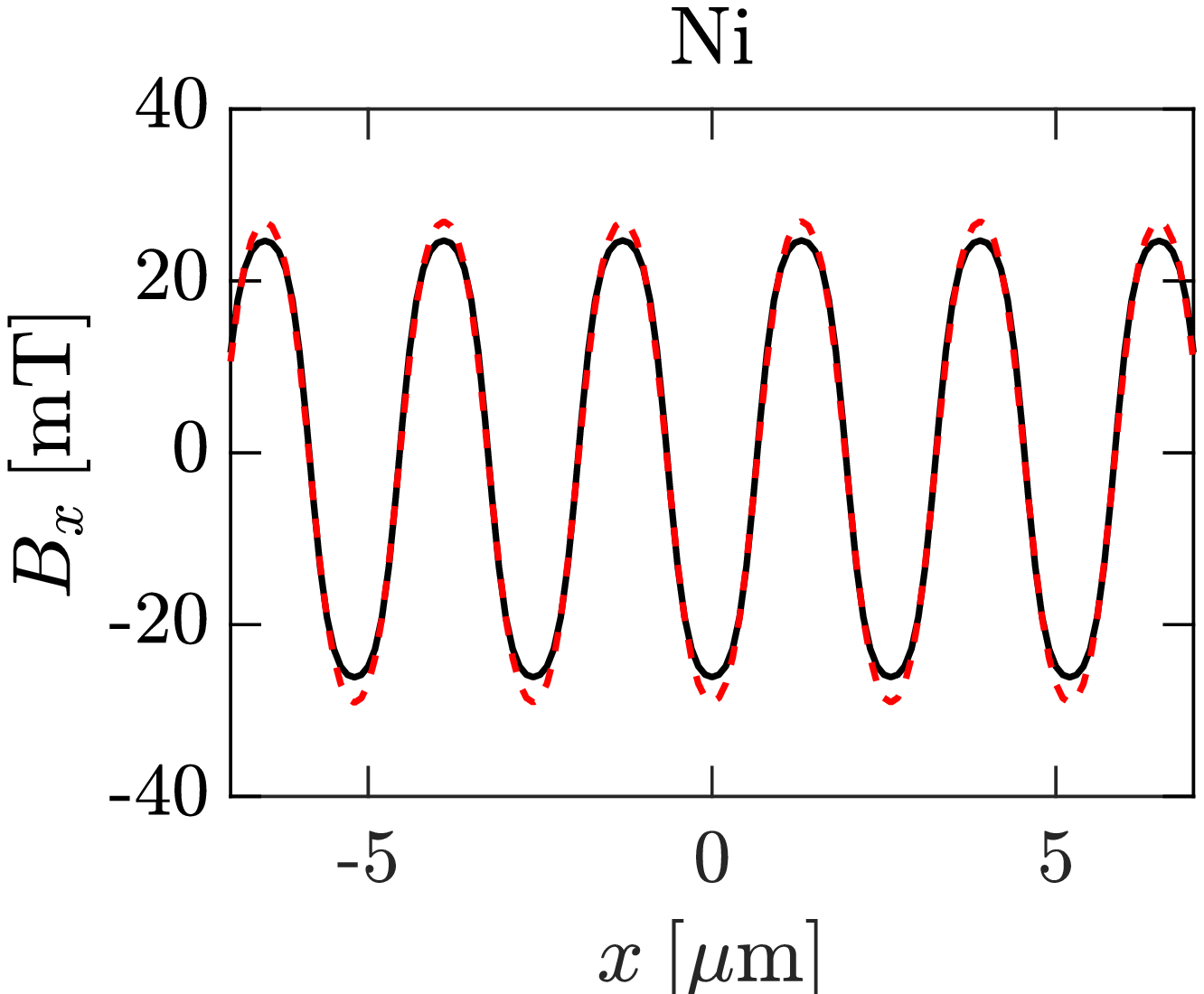}
	\caption{ Magnetic field in the 2DEG plane for Ni bars. Left panel:
		$B_z$ versus $x$. Right panel: $B_x$ versus $x$.
		The parameter sets are: black solid line $\delta z=0.2\:\mu\mathrm{m},  \ z=0.4\:\mu \mathrm{m}$,
		red dashed line $\delta z=0.3\:\mu\mathrm{m},  \ z=0.5\:\mu\mathrm{m}$.
	}
	\label{Ni}
\end{figure}
%%%%%%%%%%%%%%%%%
Plots of $B_z$ versus $x$ for Ni bars for $\delta z=0.2\:\mu \mathrm{m},  \ z=0.4\:\mu \mathrm{m}$
and for $\delta z=0.3\:\mu \mathrm{m},  \ z=0.5\:\mu \mathrm{m}$ are presented in the left panel
of Fig.~\ref{Ni}. Fields for these parameter sets are practically
indistinguishable and close to the simple sinusoidal dependence with
amplitude $30$\:m T.
In the right panel of Fig.~\ref{Ni} we plot the $x$-component of magnetic
field for the same sets of parameters. Of course the $x$-component does not
influence hydrodynamics, but it remains instructive to see the magnitude
and the $x$-dependence.

\begin{figure}[ht!]
	\includegraphics[width=0.23\textwidth]{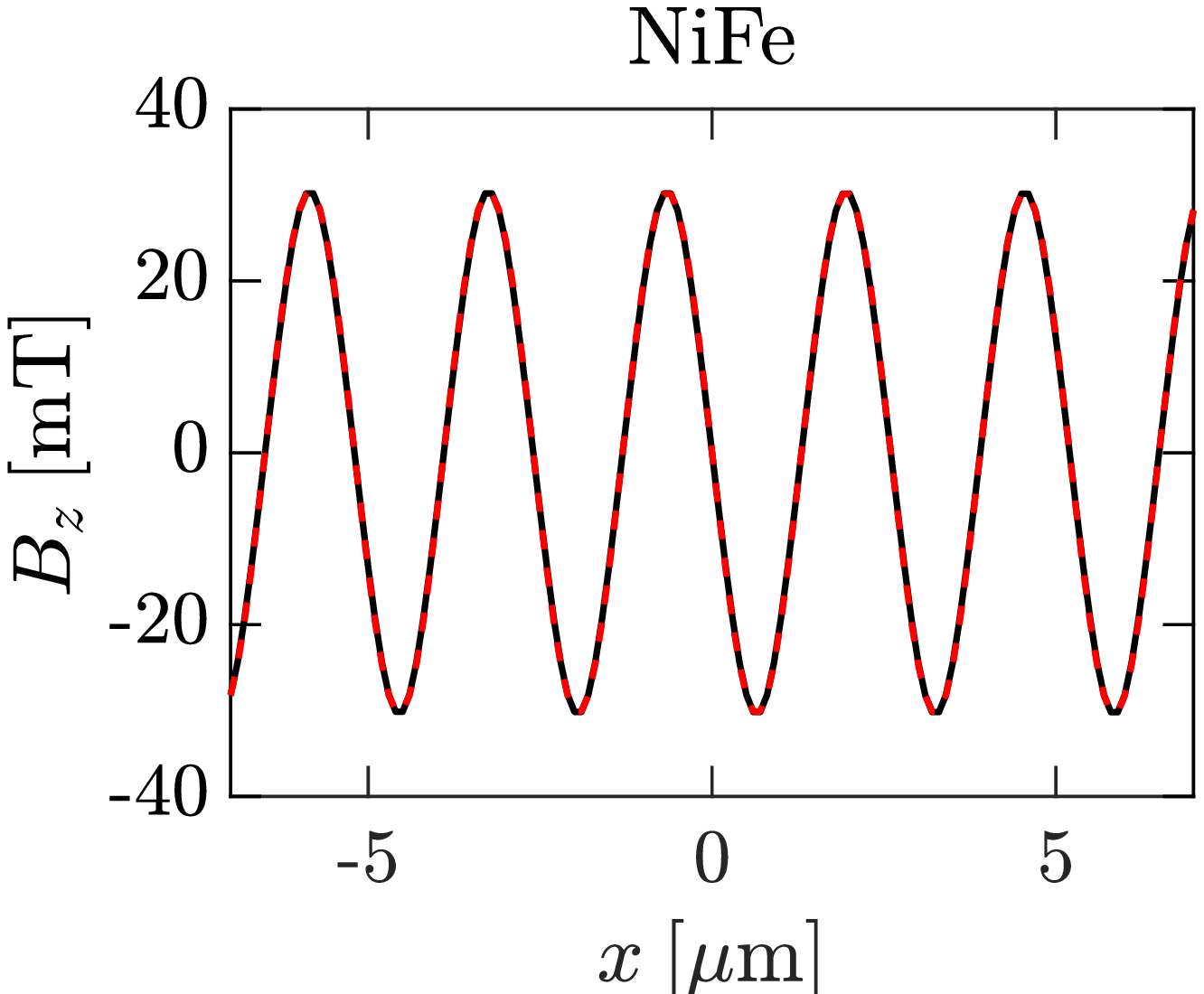}
	\includegraphics[width=0.23\textwidth]{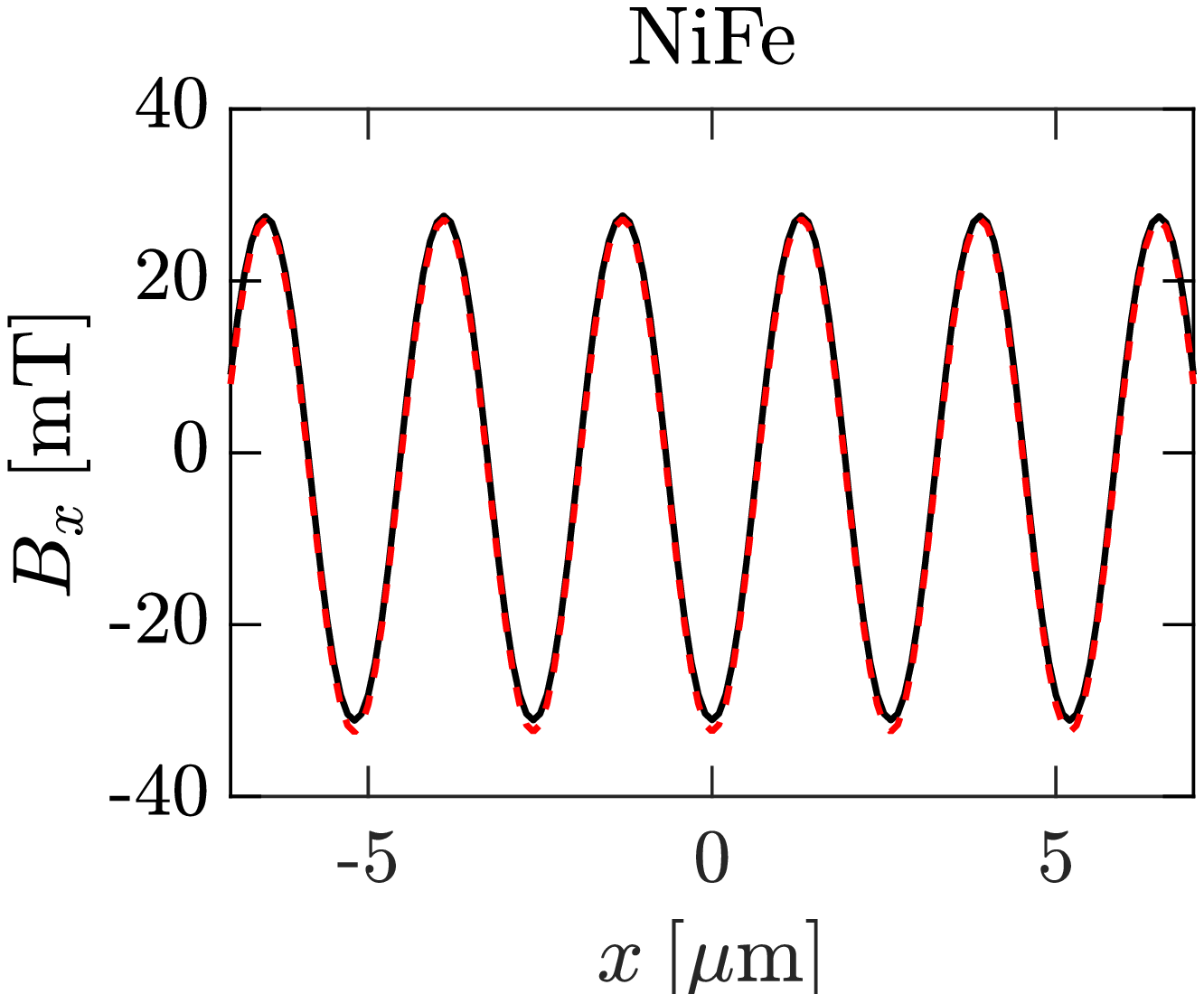}
	\caption{ Magnetic field in the 2DEG plane for NiFe bars. Left panel:
		$B_z$ versus $x$. Right panel: $B_x$ versus x.
		The parameter sets are: black solid line $\delta z=0.2\:\mu\mathrm{m},
		\ z=0.75\:\mu\mathrm{m}$, red dashed line $\delta z=0.3\mu m,  \ z=0.87\:\mu\mathrm{m}$.
	}
	\label{NiFe}
\end{figure}

Plots of $B_z$ versus $x$ for NiFe bars for ($\delta z=0.2\:\mu \mathrm{m},  \ z=0.75\:\mu \mathrm{m}$)
and for ($\delta z=0.3\:\mu \mathrm{m},  \ z=0.87\:\mu \mathrm{m}$) are presented in the left panel
of Fig.~\ref{NiFe}. Again, fields for these parameter sets are practically
indistinguishable and closely follow a sinusoidal pattern as desired. 
In the right panel of Fig.~\ref{NiFe} we plot the $x$-component of magnetic
field for the same sets of parameters.
%%%%%%%%%%%%%%%%%%%%%%%%%%%%%%%%%%%%%%%%%%%%%%%%%
%%%%%%%%%%%%%%%%%

\section{Conclusions} \label{sec:concl}
In this work we propose a new method to study viscous hydrodynamic flow of electrons in
2D mesoscopic systems. This method involves the use of micromagnets to drive oscillatory flow everywhere in the fluid domain, promoting viscous dissipation not just in the boundary layer but also in the bulk. The proposed method allows for improved precision of 
hydrodynamic measurements by at least
an order of magnitude compared to existing state-of-the-art experiments.
With such an improvement in accuracy, electron hydrodynamics presents itself as a novel and
efficient method to study electron-electron correlations.

\section*{ Acknowledgements} 

We acknowledge important discussions with Alexander Hamilton, Igor Zutic, Oleh Klochan and Daisy Wang. This work was supported by the Australian
Research Council Centre of Excellence in Future Low-
Energy Electronics Technologies (CE170100039).

\appendix
\section{Perturbative solution of boundary layer to second order} \label{app_1}

In Sec.~\ref{sec:pert_boundary_layer} we presented the perturbative boundary layer solution to first order. This first order solution was derived by solving Eq.\eqref{NS1} for the stream function, defined in Eq.~\eqref{str}, to first obtain the driven solution and then again for a judiciously crafted free solution. The generalized complete first order solution was then subject to the boundary conditions, Eq.\eqref{bc}, in order to completely define all coefficients. Now, this process can be repeated ad infinitum to obtain higher order corrections of the series
\begin{equation}
	\psi = v_0 y + v_0\sum_{n=1}^{\infty} (\omega \tau)^n  \psi^{(n)}.  
\end{equation}
Here $\psi^{(n)}$ independently satisfies the boundary conditions Eq.~\eqref{bc} at each order of correction.
%\begin{equation}
%\label{bc_n}
%    \partial_x \psi^{(n)}|_{y=0} = (\partial_y^2 - \partial_x^2)\psi^{(n)}|_{y=0} = 0. 
%\end{equation}
For simplicity we write distance in units $D = 1$. We may decompose $\psi^{(n)}$ into free and  driven components as we did to first order in the text
\begin{equation}
	\psi^{(n)} = \psi_f^{(n)} + \psi_d^{(n)},\quad  (-\nabla^2 + \nabla^4)\psi_f^{(n)} = 0.
\end{equation}
The free solution must be of the form
\begin{equation}
	\label{psi_f}
	\psi_f^{(n)} = \sum_{m=1}^{n}\sum_{s=0,1} A^{(n)}_{m,s} \cos(m g x ) e^{-\lambda_{m,s}y},
\end{equation}
where
\begin{equation}
	\label{lambdas}
	\lambda_{m,s} = \sqrt{m^2g^2 + s},\quad s=0,1.
\end{equation}

The Stokes operator, $-\nabla^2 + \nabla^4$, becomes  invertible when acting on the driven solution. Therefore the driven solution can be written as
\begin{equation}
	\psi_d^{(n)} = \sum_{\lambda} \sum_{m=1}^{n} B^{(n)}_m(\lambda) \cos(m g x ) e^{-\lambda y},
\end{equation}
%where $\lambda$ is chosen from the set of $\lambda_{m,s}$ in Eq.~\eqref{lambdas}.
where $\lambda$ is chosen from the set of $\lambda_{m',s}$ given by the formula in Eq.~\eqref{lambdas}, defined for the indices $0\leq m'<n$ and $s=0,1$.
We also note that, in this choice,  the term with $B_m^{(n)}(\lambda_{m,s})$ never appears, as it is part of the homogeneous solution
(i.e. it belongs to the kernel of the Stokes operator). 

To first order the Stokes equation, Eq.~\eqref{NS1}, reads
\begin{equation}
	\sum_{\lambda}  B^{(1)}_1(\lambda)  e^{-\lambda y} ( \lambda^2-\lambda_{1,0}^2)( \lambda^2 - \lambda^2_{1,1}) = g.  
\end{equation}
This equation is satisfied only for $\lambda = \lambda_{0,0} = 0$. Therefore the only non-vanishing coefficient is 
\begin{equation}
	B^{(1)}_1(\lambda_{0,0}) = \frac{g}{\lambda_{1,0}^2  \lambda^2_{1,1}}  = \frac{1}{g  (g^2 + 1)}.
\end{equation}
The remaining two coefficients, $A^{(1)}_{1,s}$ where $s=0,1$, may be determined from the boundary conditions, Eq.~\eqref{bc}. 
\begin{eqnarray}
	\sum_{s = 0,1} A^{(1)}_{1,s} + B_1^{(1)}(\lambda_{0,0}) &=&0,\\
	\sum_{s = 0,1} A^{(1)}_{1,s} \lambda_{1,s}^2&=&0.   
\end{eqnarray}
The solutions are $A^{(1)}_{1,s} = - g(-1)^s \lambda_{1,s}^{-2}$ where $s=0,1$, or explicitly
\begin{eqnarray}
	A_{1,0}^{(1)} &=& -\frac{g}{\lambda_{1,0}^2} =  -\frac{1}{g},\\  A_{1,1}^{(1)} &=& \frac{g}{\lambda_{1,1}^2} = \frac{g}{g^2 +1 }.
\end{eqnarray}
The first order solution is completely defined by the coefficients $A^{(1)}_{1,0},A^{(1)}_{1,1}$ and $B^{(1)}_1(\lambda_{0,0})$, written explicitly as Eq.~\eqref{1st} in the main text.

The first order solution, $\psi^{(1)}$, acts as a source for the second order driven solution, $\psi_d^{(2)}$. By solving the Stokes equation, Eq.~\eqref{NS1}, we  find the non-vanishing coefficients  of $\psi_d^{(2)}$ to be
\begin{multline}
	B_m^{(2)} (\lambda_{1,s}) = \frac{(-1)^s g^2   }{2\lambda_{1,s}(\lambda_{1,s}^2 - \lambda_{m,0}^2)(\lambda_{1,s}^2 - \lambda_{m,1}^2)},\\ \mathrm{for}\quad m = 0,2,\: s= 0,1.
\end{multline}
%\begin{multline}
%B_m^{(2)} (\lambda_{1,s}) = \frac{- g %\lambda_{1,s} A_{1,s}^{(1)} }{2(\lambda_{1,s}^2 - %\lambda_{m,0}^2)(\lambda_{1,s}^2 - %\lambda_{m,1}^2)},\\ \mathrm{for}\quad m = 0,2,\: %s= 0,1.
%\end{multline}

To satisfy the boundary conditions, Eq.~\eqref{bc}, the second order free solution has non-vanishing coefficients $A^{(2)}_{m,s}$ for  $m=0,2$ and $s=0,1$. The coefficient $A^{(2)}_{0,0}$ precedes a constant term and thus its value does not affect the current and voltage distribution. 
%remains undetermined as the velocity field is given by the spatial derivatives of the stream function. 
%\begin{eqnarray}
%    A^{(2)}_{0,0} &=& 0.
%\end{eqnarray}
%Meanwhile, the no-penetration condition gives
%\begin{equation}
%A^{(2)}_{0,1} = -\sum_{s=0,1} %\lambda_{1,s}B_0^{(2)}(\lambda_{1,s}).
%\end{equation}
%\begin{eqnarray}
%A^{(2)}_{0,1} &=& %-\frac{g^2}{2}\sum_{s=0,%1}\frac{ (-1)^s %}{\lambda_{1,s}(\lambda%_{1,s}^2 - 1)}.
%\end{eqnarray}
%Finally, from the no-stress %condition we obtain
%\begin{eqnarray}
%A_{2,0}^{(2)} &=& %\sum_{s=0,1} %(\lambda_{1,s}^2 - %\lambda_{2,1}^2) %B^{(2)}_2(\lambda_{1,s}),\\
%A_{2,1}^{(2)} &=& %\sum_{s=0,1} %(\lambda_{2,0}^2 - %\lambda_{1,s}^2) %B^{(2)}_2(\lambda_{1,s}).
%\end{eqnarray}
The boundary conditions provide three equations for the remaining three coefficients. Setting  $A^{(2)}_{0,0}$ appropriately, we have
\begin{multline}
	A_{m,r}^{(2)}  =  \frac{g^2}{2} \sum_{s=0,1}\frac{(-1)^{r+s}}{\lambda_{1,s} (\lambda_{1,s}^2 - \lambda_{m,r}^2)},\\ \mathrm{for}\quad m = 0,2,\: r= 0,1.
\end{multline}

Evidently, the leading term of the second order correction decays at a faster rate than that of the first order correction. Higher order boundary layers decay  at a rate faster than  that of $\exp(- g y)$ or $\exp(-y/D)$. Thus in the perturbative regime, $\omega\tau < 1$, the width of the boundary layer is finite with an effective width comparable to the largest of $D$ and $1/g$. 

\bibliography{electron_fluid_refs_Aug_19}
\end{document}